\documentstyle[epsf,axodraw,iopconfJE]{article}

\newcommand{\be}{\begin{equation}}
\newcommand{\ee}{\end{equation}}

\newcommand{\bea}{\begin{eqnarray}}
\newcommand{\nn}{\nonumber}
\newcommand{\eea}{\end{eqnarray}}

\newcommand{\Ce}{C_\epsilon}
\newcommand{\De}{D_\epsilon}

\def\gsim{\mathrel{\rlap {\raise.5ex\hbox{$ > $}}
{\lower.5ex\hbox{$\sim$}}}}
\def\lsim{\mathrel{\rlap {\raise.5ex\hbox{$ < $}}
{\lower.5ex\hbox{$\sim$}}}}

\def\gappeq{\mathrel{\rlap {\raise.5ex\hbox{$>$}}
{\lower.5ex\hbox{$\sim$}}}}

\def\lappeq{\mathrel{\rlap{\raise.5ex\hbox{$<$}}
{\lower.5ex\hbox{$\sim$}}}}

\begin{document}
\title{Probing Models of Quantum Space--Time Foam}

\author{John  Ellis~\footnote{E-mail:
John.Ellis@cern.ch}}

\affil{CERN, Theory Division, Geneva 23 CH-1211, Switzerland}

\author{N.E. Mavromatos~\footnote{E-mail:
Nikolaos.Mavromatos@cern.ch}}

\affil{Wheatstone Physics Laboratory, King's College London,
Strand, London WC2R 2LS, UK}

\author{D.V. Nanopoulos~\footnote{E-mail:
dimitri@soda.physics.tamu.edu}}

\affil{Department of Physics, 
Texas A \& M University, College Station, 
TX~77843-4242, USA, \\
Astroparticle Physics Group, Houston
Advanced Research Center (HARC), Mitchell Campus,
Woodlands, TX 77381, USA, and
Academy of Athens,
Chair of Theoretical Physics, 
Division of Natural Sciences, 28~Panepistimiou Avenue, 
Athens 10679, Greece.}

\beginabstract
We review the possibility that quantum fluctuations
in the structure of space-time at the Planck scale
might be subject to experimental probes. We study
the effects of space-time foam in an approach inspired by
string theory, in which
solitonic $D$-brane
excitations 
are
taken into account when
considering the ground state. We model the
properties of this medium by analyzing
the recoil of a $D$ particle which is induced by the
scattering of a closed-string state. We find that
this recoil causes an energy-dependent perturbation
of the background metric, which in turn induces an
energy-dependent refractive index {\it in vacuo}, and
stochastic fluctuations of the light cone. We show how
distant astrophysical sources such as Gamma-Ray Bursters (GRBs)
may be used to test this possibility, making an
illustrative analysis of GRBs whose redshifts have
been measured. Within this framework, we also discuss the propagation
of massive particles and the possible appearance of
cosmological vacuum energy that relaxes towards zero.
We also discuss $D$-brane recoil in models with
`large' extra dimensions.
\endabstract

\begin{center}
ACT-10/99~~~~~~~~~~~~CTP-TAMU-40/99~~~~~~~~~~~~gr-qc/9909085
\end{center}

\section{Introduction and Orientation}

The concept of space time foam is an old one,
first suggested by J.A. Wheeler~\cite{wheeler},
which has subsequently reappeared in various
forms: see, for example~\cite{foam,ellis84,ellis92,
amelino98,garay98,ashtekar99,amelino99}.
The basic intuition is that quantum-gravitational fluctuations 
in the fabric of space-time cause the vacuum to behave like
a stochastic medium. Most physicists who have studied the
problem would surely agree that
quantum-gravitational interactions must
alter dramatically 
our classical perception of the space-time continuum
when one attains 
Planckian energy scales $E \sim M_P \simeq 10^{19}$~GeV or 
Planckian distance scales $\ell \sim \ell_P \simeq 10^{-33}$~cm,
which are
the scales where gravitational interactions 
are expected to become strong. At issue are the following questions:
how may classical space-time be altered at these scales? and is there
any way of testing these possibilities?

At first sight, it might seem impossible to 
test such a suggestion 
within the foreseeable future,
given the restrictions on the energies attainable with
particle accelerators, and their corresponding limitations as
microscopes. However, there are many instances in which new
physics has revealed itself as a novel phenomenon far below its
intrinsic mass scale $M$, a prime example being the weak interaction.
In general, new-physics are suppressed by some inverse power of 
the heavy mass scale $M$,
e.g., weak-interaction amplitudes are suppressed by $1/M_W^2$. 
However, some new-physics effects may be suppressed by just one
power of the heavy mass scale, e.g., proton-decay amplitudes in
some supersymmetric GUTs are $\propto 1/M_{GUT}$. Although
gravitational amplitudes are generally suppressed by $1 / M_P^2$,
one should be open to the possibility that some quantum-gravitational
effects might be suppressed simply by $1 / M_P$. Moreover, there are
many suggestions nowadays that $M_P$ might not be a fundamental
mass scale, and the quantum-gravitational effects might appear at
some much lower scale related to the size(s) of one or more extra
dimensions~\cite{dimopo98}.

It has sometimes been suggested that
Lorentz invariance might require
quantum-gravitational effects to
be suppressed by at least
${\cal O}[\left(E/M_P\right)^2]$,
where $E$ is a typical low-energy scale. 
However, Lorentz invariance is a casualty of
many approaches to quantum gravity, and it is
not clear how the very concept of space-time
foam could be formulated Lorentz-invariantly.
For example, many approaches to the physics of very small distances
suggest that the classical space-time continuum
may no longer exist, but might instead be replaced
by a cellular structure.

String theory, plausibly in its current formulation as $M$~theory,
is at present the best (only?) candidate for a true quantum theory
of gravity, so it is natural to ask what guidance it may offer us
into the possible observability of quantum-gravitational effects.
A first tool for this task was provided by two-dimensional string
models~\cite{emn}, but a much more powerful tool has now been provided by
$D$(irichlet branes~\cite{dbranes}. In this talk we review
one particular $D$-brane approach to the 
modelling of space-time foam~\cite{emndbrane}. A characteristic feature of
this
approach is a treatment of $D$-brane recoil effects, in an attempt
to incorporate the back-reaction of propagating particles on the
ambient metric. This leads to the sacrifice of
Lorentz invariance at the Planck scale,
and suggests that space-time-foam effects
arise already at the order ${\cal O}\left(E/M_P\right)$,
in which case they might well be observable.

We have argued in the past that
such minimally-suppressed quantum-gravitational effects
could be probed in the neutral-kaon system~\cite{ellis84}, which is
well known to be a sensitive laboratory for testing
quantum mechanics and fundamental symmetries. In this
talk, we focus more on the possibility that
the propagation of other particles such as
photons or neutrinos might be affected in a way that 
could be testable
in the foreseeable future, for instance through astrophysical
observations of pulsed sources such as
$\gamma$-ray bursters (GRBs), active galactic nuclei
(AGNs) or pulsars~\cite{amelino98}.
Our basic suggestion is that space-time foam may act as a non-trivial
optical medium with, e.g., an energy- or frequency-dependent refractive
index, through which the propagation of energetic particles might
be slowed down so that they travel at less than the speed of light $c$:
$\delta c / c \sim - E / M_P$. A secondary effect might be a
stochastic spread in the velocities of particles with identical energies.

We describe in Section 2 below the motivations for such suggestions,
starting from a (relatively) simplified 
non-technical description of the string-inspired
prototype model of space-time foam based on one
particular treatment of $D$ branes~\cite{emndbrane,ellis99}.
This model involves naturally the 
breaking of Lorentz covariance, as a 
sort of spontaneous breaking.   
The basic idea may be summarized as follows.
In the modern view of string theory, $D$ particles must be included
in the consistent formulation of the ground-state vacuum
configuration~\cite{dbranes}~\footnote{For 
stability reasons, the ground state must be 
supersymmetric. However, 
supersymmetry is not essential for our analysis, and 
we usually do not mention it explicitly until
some comments near the end of the talk.}.  
We consider a closed-string state propagating in a 
$(D+1)$-dimensional space-time, which impacts
a very massive $D$(irichlet) particle embedded in this space-time.
We argue that the scattering of the closed-string state on the 
$D$ particle induces recoil of the latter, which
distorts the surrounding space-time 
in a stochastic manner. From the point of view of the
closed-string particle and any
low-energy spectators, this is a non-equilibrium process, in which
information is `lost' during the recoil,
being carried by recoiling $D$-brane degrees of freedom that are
inaccessible to a low-energy observer. 
Thus, although the entire process is consistent with 
quantum-mechanical
unitarity, the low-energy effective theory is
characterized by information loss and 
entropy production. From a string-theory point of view,
the loss of information is encoded in a
deviation from conformal invariance 
of the relevant world-sheet $\sigma$ model that
describes the recoil, that is compensated by
the introduction of a Liouville field~\cite{distler89}, which in
turn is identified with the target time
in the approach~\cite{emn} adopted here. 

The reader who wishes may skip the next section and proceed
directly to Section 3, where we discuss the possible 
phenomenology of this approach to space-time foam. 
We discuss the appearance of an energy-dependent
refractive index {\it in vacuo} and stochastic
fluctuations in the velocities of photons of the same energy.
We then discuss the corresponding modification
of Maxwell's equations, and their consequences for the
propagation of photon pulses. Subsequently, we make a sample
analysis of data from 
distant astrophysical sources, using GRBs with measured
redshifts. Section 4 contains a discussion of the propagation of massive
pasrticles, brief comments on
the possible appearance of cosmological vacuum energy
within this approach, and an exploration of $D$-brane
recoil effects in models with `large' extra dimensions.

\section{$D$-Particle Recoil Model for Space-Time Foam} 

We recall that $D$ branes are solitons in string theory~\cite{dbranes},
of mass $m \sqrt{\alpha '} \sim 1/g_s$, where $\alpha ' \equiv \ell_s ^2$ is 
the Regge slope (string scale), and $g_s$ is the string coupling,
which is assumed throughout this work to be weak.
We consider the situation depicted in Fig.~1, namely the scattering 
of a closed-string state (which might represent a photon) on a $D$ brane
embedded in a $D + 1$-dimensional space-time. 
We assume that the $D$ brane is so heavy
that a non-relativistic approximation is sufficient. 
As already mentioned, we assume that the vacuum is
(approximately) supersymmetric, and that the vacuum energy
vanishes (approximately). We believe that
an instructive analogy may be drawn between the
closed-string/${D}$-particle system
and a system of valence electrons moving freely through an ion lattice
in a solid. Ion-lattice vibrations (c.f., the recoil of ${
D}$ particles) are well-known to   
induce effective phonon interactions, which bind the
electrons resulting in BCS superconductivity for the ground state
of this many-body system.
In this example, the physical excitations above this ground state
are quasiparticles, which are not the ordinary electrons, but
are `dressed' by their interactions with the excitations in
the background medium, affecting their properties. In the
language of relativistic cosmology, the $D$ branes can be
regarded as providing a `material reference system' (MRS).
At a fundamental level, the non-rigid recoil of $D$ branes
implies that the vacuum becomes a non-trivial medium.

\begin{figure}[htb]
\epsfxsize=2in
\centerline{\epsffile{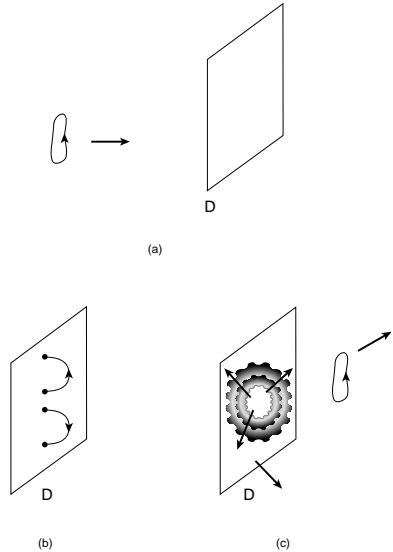}}
\vspace{0.2cm}
\caption[]{ \it The scattering of a low-energy
closed-string state on a $D$ brane: (a) the asymptotic past, 
(b) the time of impact ($X^0=0$), with 
trapping of the string state on the $D$-brane surface by 
a split into two open-string excitations, and 
(c) the asymptotic future, after two open strings recombine to emit a 
closed-string state, while the 
$D$ brane recoils with finite velocity and its internal 
state fluctuates.}
\end{figure}

The study of $D$-brane dynamics 
has been made possible by Polchinski's realization~\cite{dbranes} 
that such solitonic string backgrounds
can be described in a conformally-invariant way,
in terms of world sheets
with boundaries (thus incorporating 
open strings), on which Dirichlet boundary conditions 
for the collective target-space coordinates of the soliton are imposed.
The remarkably simple construction of Polchinski~\cite{dbranes}
opens the way to a $\sigma$-model description of 
such $D$-brane excitations, in which
the critical world-sheet string action 
is perturbed using appropriate 
boundary terms. We recall the form of the world-sheet
boundary operators describing
the excitation of a $D$ brane~\cite{dbranes}: 
\be
   {\cal V}_D = \int _{\partial \Sigma}( y_i \partial _n X^i
+ u_i X^0 \partial_n X^i)
\label{dbraneop}
\ee
where $n$ denotes the normal derivative on the boundary of the world
sheet $\partial \Sigma$,
which has at tree level the topology of a disk of size $L$, and the
$X^i~,i=1,\dots $ denote the collective excitations of the $D$ brane,
which satisfy Dirichlet boundary conditions
on the world-sheet boundary:
\be
 X^i ({\rm boundary}) = 0,~i=1,\dots,
\label{fiveb}
\ee
whilst  $X^0$ is the target time variable which satisfies  
standard Neumann
boundary conditions: $\partial _n X^0 ({\rm boundary}) = 0$.
For simplicity, we consider later in more detail the case of 
a $0$ brane, or $D$ particle~\footnote{See~\cite{EMW}
for a more general discussion of $D$-brane/$D$-brane scattering.}, with
the quantity $u_i $ in (\ref{dbraneop}) denoting its
velocity, and $y_i$ its initial position. In this case, the
operators (\ref{dbraneop}) describe shifts and
motion of the $0$ brane, and so can be
thought of as generating the action of the Poincar\'e group on the 
$D$ particle, with the $y_i$ parametrizing translations and the $u_i$
parametrizing boosts. In the general $D$-brane case, these
represent translations and boosts acting on the surface ${\cal S}$
of the $D$ brane.

As a first step
towards the quantum theory of the scattering 
of a closed-string state on such a $D$ brane,
we consider its
motion towards the (initially fixed) surface 
${\cal S}$ shown in Fig.~1(a), the latter viewed as a string 
soliton background.
In the general $D$-brane case, the surface ${\cal S}$ divides
the target space-time into two regions. 
The closed-string state 
is initially far away from the 
surface of the brane. At a certain moment,
say $X^0=0$, 
the incoming closed-string state finds itself lying partly outside and
partly inside the $D$-brane surface. There are then two possibilities
to be considered: it may be either absorbed Fig.~1(b) or 
rescattered as in Fig.~1(c).
In general, quantum scattering
on the $D$ brane excites an open-string state on its surface,
which in the scattering case of Fig.~1(c)
also emits another closed-string state. The quantum excitation and
emission processes are both described by closed-to-open
string amplitudes, which are non-zero
in a world-sheet theory with boundaries. The open-string states
are excitations on the $D$-brane collective-coordinate surface. 
As was shown in~\cite{cardy}, such processes can be described
in terms of data of the bulk theory. As we showed
in \cite{emndbrane}, 
tracing over such 
excitations results in a quantum modification of the 
Hawking-Bekenstein area law for the entropy, which
has been shown to hold in tree-level treatments of $D$ branes.

This modification is due to the
essence of our approach, which reflects the basic property 
of General Relativity that {\it there are no rigid 
bodies}, and hence the recoil fluctuations of the brane
and their back-reaction on
the surrounding space-time medium cannot be neglected. 
 
A correct world-sheet 
treatment of recoil requires~\cite{kogan96}
an operator with non-zero matrix
elements between different $D$- (in our case $0$-)brane states. 
This can be achieved in the impulse approximation 
by introducing a Heaviside-function factor $\Theta (X^0)$
into the second operator in (\ref{dbraneop}), so as to
describe a $0$ brane that starts moving at
time  $X^0=0$, when the initial position of the
$0$ brane at $X^0=0$ is given by the $y_i$. 
To determine the precise form of 
the recoil operator in our case, we observe that the leading quantum
correction to the scattering of a closed string
on a $D$ brane is given by an annulus, as shown in Fig.~2(a). This
is divergent in the limit where the annulus is pinched, as shown in
Fig.~2(b). The 
weakly-coupled string limit: $g_s \rightarrow 0$, that we consider here, 
corresponds in leading order to a one-open-string-loop (annulus) analysis
for a 
semi-classical (heavy) $0$ brane ($D$ particle), since the mass of the 
latter is $M_D \propto 1/g_s$ in natural string units.

\begin{figure}[htb]
\epsfxsize=2in
\centerline{\epsffile{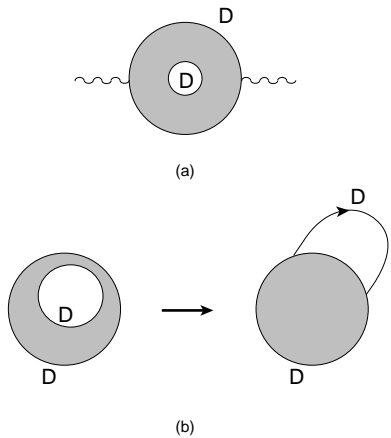}}
\vspace{0.2cm}
\caption[] { \it (a) : World-sheet
annulus diagram for the leading quantum correction to the 
propagation of a string state in a $D$-brane background, and 
(b) the pinched annulus configuration which is the dominant divergent 
contribution to the quantum recoil.}
\end{figure}

In the case of a $D$-brane string soliton, its recoil after
interaction with a closed-string state~\cite{kogan96}
is characterized by a pair of logarithmic operators~\cite{lcft}:
\be
C_\epsilon \sim \epsilon \Theta_\epsilon (t),\qquad
D_\epsilon \sim t \Theta_\epsilon (t)
\label{logpair}
\ee
defined on the boundary $\partial \Sigma$ of the string
world sheet. The operators (\ref{logpair}) act
as deformations of the conformal field theory on the
world sheet: $u_i \int _{\partial \Sigma} \partial_n X^i \De$ 
describes the shift of the $D$
brane induced by the scattering, where
$u_i$ is its recoil velocity, and $ y_i 
\int _{\partial \Sigma} \partial_n X^i \Ce$
describes quantum fluctuations in the initial
position $y_i$ of the $D$ particle. It has been
shown~\cite{mavro+szabo} that
energy-momentum is conserved during the recoil process:
\be
u_i = k_1 - k_2, 
\label{cons}
\ee
where $k_1 (k_2)$ is the momentum of 
the propagating closed-string state before (after) the recoil,
as a result of the summation over world-sheet genera.
Thus the result (\ref{cons}) is an exact result, as far as 
world-sheet perturbation theory is concerned. 
We also note that $u_i = g_s P_i$, where $P_i$ 
is the momentum and $g_s$ is the string coupling,
which is assumed here to be weak enough
to ensure that $D$ 
branes are very massive, with mass $M_D=1/(\ell _s g_s)$,
where $\ell _s$ is the string length.   

The correct specification of the logarithmic pair (\ref{logpair})
entails a regulating 
parameter $\epsilon \rightarrow 0^+$, which
appears inside the $\Theta_\epsilon (t)$ operator:
$\Theta_\epsilon (t) = \int \frac{d\omega}{2\pi}\frac{1}{\omega
-i\epsilon} 
e^{i\omega t}  $. In order to realize
the logarithmic algebra between the operators $C$ and $D$,
one takes~\cite{kogan96}:
\be
\epsilon^{-2} \sim {\rm Log}\Lambda/a \equiv \alpha
\label{defeps}
\ee
where $\Lambda$ ($a$) are infrared (ultraviolet) world-sheet cut-offs. 
The pertinent two-point functions then have the following 
form~\cite{kogan96}: 
\bea   
<C_{\epsilon}(z) C_{\epsilon}(0) > &\sim& 0 + O[\epsilon^2] \nn \\
   <C_{\epsilon}(z) D_{\epsilon}(0) > &\sim & 1 \nn \\
    <D_{\epsilon}(z)D_{\epsilon}(0)> &\sim& 
\frac{1}{\epsilon ^2} -2 \eta \log|z/L|^2
\label{twopoint}
\eea
up to an overall normalization factor, 
which is the logarithmic algebra~\cite{lcft} in 
the limit $\epsilon \rightarrow 0^+$,
modulo the leading divergence in the $<\De\De>$ recoil correlator.
This leading divergent term 
will be important for our subsequent analysis.

The recoil deformations of the $D0$ brane (\ref{logpair})
are relevant deformations, in the sense of conformal field theory, with
anomalous dimension  $-\epsilon^2/2 $. 
However~\cite{mavro+szabo}, the velocity operator 
$\De$ (\ref{logpair}) becomes exactly marginal,
in a world-sheet renormalization-group sense, when it is divided
by $\epsilon $, in which case the recoil
velocity is renormalized~\cite{mavro+szabo}:
\be
u_i \rightarrow {\overline u}_i \equiv u_i /\epsilon 
\label{marginal}
\ee
and becomes exactly marginal, playing the r\^ole 
of the physical velocity of the recoiling $D$ particle. 

Although such a renormalization 
is compatible with global world-sheet scaling, 
local world-sheet scale (conformal) symmetry 
is broken by 
the non-marginal character of the deformations
(\ref{logpair}), and
restoration of conformal 
invariance requires Liouville dressing~\cite{distler89}. 
To determine the effect of such dressing on the space-time 
geometry, it is essential  to write~\cite{emndbrane,ellis99} the 
boundary recoil deformation
as a bulk world-sheet deformation
\begin{equation}
    \int _{\partial \Sigma} d\tau {\overline u}_{i} X^0 \Theta_\epsilon
(X^0) \partial_n X^i =
\int _\Sigma d^2\sigma \partial_\alpha \left([{\overline u}_{i}
X^0]\Theta_\epsilon
(X^0) \partial_\alpha X^i \right) 
\label{a1}
\end{equation}
where the ${\overline u}_{i}$ denote the 
renormalized
recoil couplings (\ref{marginal}), 
in the sense discussed in~\cite{mavro+szabo}.
As we have already mentioned, the couplings (\ref{a1}) are marginal on a
flat world sheet, and become marginal on a curved  
world sheet if one 
dresses~\cite{distler89}  the bulk integrand with
a factor $e^{\alpha_{i}\phi}$, where $\phi$ is the Liouville
field
and $\alpha_{i}$ is the gravitational conformal dimension. This 
is related to the flat-world-sheet anomalous dimension $-\epsilon^2/2$ of
the recoil operator, 
viewed as a bulk world-sheet deformation, as follows~\cite{distler89}:
\begin{equation}
\alpha_{i}=-\frac{Q_b}{2} + 
\sqrt{\frac {Q_b^2}{4} + \frac {\epsilon^2}{2} } 
\label{anom}
\end{equation}
where $Q_b$ is the central-charge deficit of the bulk world-sheet theory.
In the recoil problem at hand, as discussed in~\cite{kanti98},
$Q_b^2 \sim \epsilon^4/g_s^2$ for weak deformations.
This yields $\alpha_{i} \sim -\epsilon $ to leading order in 
perturbation theory in $\epsilon$, to which we restrict ourselves here. 

We next remark that, as the analysis of~\cite{emndbrane} indicates, 
the $X^0$-dependent operator
$\Theta_\epsilon (X^0)$ scales as follows with $\epsilon$ 
for $X^0 >0$:  
$\Theta_\epsilon(X^0) \sim e^{-\epsilon X^0}
\Theta(X^0)$, where $\Theta(X^0)$ is a Heaviside step function
without any field content, evaluated in the limit $\epsilon \rightarrow 0^+$. 
The bulk deformations, therefore, yield the following 
$\sigma$-model terms:
\begin{equation} 
\frac{1}{4\pi \ell_s^2} \epsilon \sum_{I=m+1}^{D-1} g_{Ii} X^I 
e^{\epsilon(\phi_{(0)} - X^I_{(0)})}\Theta(X^I_{(0)})~\int _\Sigma
\partial^\alpha X^I 
\partial^\alpha y_i~
\label{bulksigma}
\end{equation} 
where the subscripts $(0)$ denote world-sheet zero modes.

When we interpret the Liouville zero mode $\phi_{(0)}$ as target
time, $\phi_{(0)} \equiv X^0 = t$, 
the deformations (\ref{bulksigma}) yield 
space-time metric deformations in 
a $\sigma$-model sense, which were interpreted in~\cite{emndbrane}
as expressing the distortion of the space-time 
surrounding the recoiling $D$-brane soliton. 
For clarity,
we now drop the subscripts $(0)$ for the rest of this paper.
The resulting space-time distortion is then
described by the metric elements:  
\be
G_{ij} =\delta _{ij} , 
G_{00}=-1, G_{0i}=\epsilon(\epsilon y_i + 
\epsilon {\overline u}_i t)\Theta _\epsilon (t),
~i=1, \dots D-1 
\label{metric}
\ee
where the suffix $0$ denotes temporal (Liouville) components. 

The presence of $\Theta (t)$ functions 
in (\ref{metric}) implies that 
the induced space-time is piecewise continuous~\footnote{This has
important implications for non-thermal particle production
and decoherence for a spectator low-energy field theory
in such space-times, as discussed in~\cite{kanti98,emndbrane}.},
with a singularity
in the Riemann curvature scalar as $t \rightarrow 0$:
\begin{eqnarray}
R_{\mu\nu\rho\sigma}\,R^{\mu\nu\rho\sigma}&=& 
4\,(D-1)\,\epsilon^4\,[\delta_{\epsilon} (t)]^2 + {\cal O}(\epsilon^6)
\nn \\[3mm]
R_{\mu\nu}\,R^{\mu\nu}&=&
D\,(D-1)\,\epsilon^4\,[\delta_{\epsilon} (t)]^2+ {\cal O}(\epsilon^6)
\nn \\[3mm]
R^2 &=& 4\,(D-1)^2\,\epsilon^4\,[\delta_{\epsilon} (t)]^2
+{\cal O}(\epsilon^6)
\label{repsilon}
\end{eqnarray}
where $\delta_{\epsilon}(t)$ is the appropriate derivative
of $\Theta_{\epsilon}(t)$~\cite{kanti98}.
The reader should not be alarmed by the appearance of the
$[\delta_{\epsilon}(t)]^2$ factors, which do not make 
$\epsilon$-dependent divergent
contributions to the physical quantities of interest to us, such as the
integrated central-charge
deficit $Q$, etc., even in the limit $\epsilon \rightarrow 0^+$.
In our regularization, it can be easily shown that in this
limit $\Theta_{\epsilon}(0) \rightarrow \pi$ whilst
$\delta_{\epsilon}(0)$ becomes formally a linearly-divergent
integral that is independent of $\epsilon$.

We note that the derivation of the space-time discussed above was
essentially non-relativistic, since we worked in the approximation 
of a very heavy $D$ brane. This is the reason why the singularity
in the geometry (\ref{repsilon}) is space-like, and why
the consequent change in the quantum state also occurs on a
space-like surface.
This apparent non-causality is merely an artefact of our approximation.
We expect that, in a fully relativistic $D$-brane approach,
the space-time singularity would travel along a light-cone, and that
the quantum state would also change causally.  
However, here we are interested in the difference between early- and
late-time quantum states, and this apparent non-causal behaviour 
is irrelevant.

As shown in \cite{kanti98},
the metric (\ref{metric}) can be derived
using the equations of motion for a suitable 
effective action $S$ induced by quantum $D$-brane effects:
\begin{equation}
S=\int d^Dx \sqrt{-G}\,\alpha' e^{-2\Phi}\,\hat{R}_{GB}^2
\label{action}
\end{equation}
where $\Phi = Q t$ is a linear dilaton field~\cite{aben},
with $Q$ the appropriate central charge deficit of the 
world-sheet (non)conformal theory associated with the 
metric deformation (\ref{metric}), which was 
calculated in \cite{kanti98},   
and $\hat{R}_{GB}^2$ is
the ghost-free Gauss-Bonnet quadratic combination of the
Riemann tensor, Ricci tensor and curvature scalar:
\begin{equation}
\hat{R}_{GB}^2 = R_{\mu\nu\rho\sigma}\,
R^{\mu\nu\rho\sigma}-4 R_{\mu\nu}\,R^{\mu\nu}+R^2
\label{gaussbonnet}
\end{equation}
This action should not be confused with the ordinary Einstein
action describing space-time far away from the defect.
Our point is rather to show that the 
target-space-time deformations 
induced by recoil of the $D$--particle 
are compatible with equations of motion obtained from 
an effective action, and hence can be considered as 
consistent string backgrounds on which 
ordinary low-energy matter propagates. 
The next Section 
will consider the physical consequences for such 
propagation. 

Before that, however, it is essential 
to discuss in some detail renormalization-group rescaling
on the world sheet, with the aim of establishing a connection
between the target-time variable $t$ and $\epsilon$. 
To this end, we consider a finite-size world-sheet 
scale transformation
\be
L  \rightarrow L' = L e^{t}
\label{fsscaling}
\ee 
where $L$ is the size of the world sheet 
in units of the ultraviolet cut-off:   
this is the only type 
of dimensionless scale that makes
physical sense for the open-string world sheet. 
The relation (\ref{defeps}) between
$\epsilon$ and $L$ entails the following transformation
of $\epsilon$:
\be
\epsilon^2  \rightarrow \epsilon'^2 =
 \frac{\epsilon^2}{1 + 4\eta \epsilon^2 t}
\label{epsilontransform}
\ee
We deduce from the scale dependences of the correlation functions
(\ref{twopoint}) that the corresponding transformations of
$ C_{\epsilon}$ and $D_{\epsilon}$ are:
\bea
D_{\epsilon} &\rightarrow& D_{\epsilon'} =
D_{\epsilon} - t C_{\epsilon} \nn \\
C_{\epsilon} &\rightarrow& C_{\epsilon'}= C_{\epsilon}
\label{correspond}
\eea
We emphasize that this transformation law is
unambiguous.

The corresponding transformation laws for
the couplings $y_i$ and $u_i$, which are conjugate to 
$ D_{\epsilon}$ and $C_{\epsilon}$,  are
\be
 u_i \rightarrow u_i~~,~~y_i \rightarrow y_i + u_i t 
\label{scale2}
\ee
These are consistent with the interpretations of $u_i$
as the velocity after the scattering process
and $y_i$ as the spatial collective coordinates 
of the brane, if and only if
the parameter 
$\epsilon^{-2}$ is
identified with the {target Minkowski time} $t$ 
for $t \gg 0$ after the collision:
\be
\epsilon ^{-2} \simeq t
\label{epstime}
\ee
We have assumed in this analysis
that the velocity $u_i$ is small, as is
appropriate in the weak-coupling r\'egime studied 
here. The $D$-brane $\sigma$-model formalism
is completely relativistic, and we anticipate that a complete
treatment beyond the one-loop order discussed here will
incorporate correctly all relativistic effects, including
Lorentz factors wherever appropriate.

In view of (\ref{epstime}), one observes that for $t \gg 0$ 
the metric (\ref{metric}) becomes to leading order:
\be
G_{ij} =\delta _{ij}, 
G_{00}=-1, G_{0i} \sim  {\overline u}_i, \qquad ~i=1, \dots D-1    
\label{metric2} 
\ee
which is constant in space-time. 
However, the effective metric depends on the energy content 
of the low-energy particle that scattered
on the D--particle, because of
momentum conservation during the recoil process
(\ref{cons}). This energy dependence is the 
primary deviation from Lorentz invariance 
induced by the $D$--particle recoil.

\section{Phenomenological Implications for the Propagation of
Photons} 

\subsection{Refractive Index {\it in Vacuo}} 

We now proceed to discuss possible 
phenomenological consequences of the above phenomena,
starting with the propagation of photons and
relativistic particles. The above discussion of
recoil suggests that the space-time background should be regarded
as a non-trivial.
Light propagating through media with
non-trivial optical properties may
exhibit a frequency-dependent refractive
index, namely a variation in the light
velocity with photon energy. Another possibility is
a difference between the velocities of light with
different polarizations, namely birefringence,
and a third is a diffusive spread in the apparent velocity of
light for light of fixed energy (frequency). Within the
framework described in the previous Section,
the first~\cite{amelino98} and third~\cite{ellis99} 
effects have been dervied via a
formal
approach based on a Born-Infeld Lagrangian using
$D$-brane technology~\footnote{The possibility 
of birefringence has been raised~\cite{pullin} within a
canonical approach to quantum gravity, but we do not pursue such a
possibility here.}. A different approach
to light propagation has been taken in~\cite{ford}, where
quantum-gravitational fluctuations in the light-cone have been
calculated. Here we use this formalism together with
the microscopic model background obtained in the previous Section
to derive a non-trivial refractive index and a diffusive spread
in the arrival times of photons of given frequency. 

We first review briefly the analysis in \cite{ford},
which considered gravitons in a squeezed coherent state, the natural result 
of quantum creation in the presence of black holes.  
Such gravitons induce quantum fluctuations in the space-time metric, 
in particular fluctuations in the light-cone~\cite{ford},
i.e., stochastic fluctuations in the 
velocity of light propagating through this `medium',
Following~\cite{ford},
we consider a flat background space-time 
with a linearized perturbation, corresponding to the 
invariant metric element
$ds^2=g_{\mu\nu}dx^\mu dx^\nu = 
\left(\eta_{\mu\nu} + h_{\mu\nu}\right)dx^\mu dx^\nu 
= dt^2 - d{\overline x}^2 + h_{\mu\nu}dx^\mu dx^\nu $.
Let 
$2 \sigma (x,x')$ be the squared geodesic separation 
for any pair 
of space-time 
points $x$ and $x'$, and let $2 \sigma_0(x,x')$ denote the corresponding 
quantity in a flat space-time background. 
In the case of small 
gravitational perturbations about the flat background,
one may expand
$\sigma = \sigma _0 + \sigma_1 + \sigma_2 + \dots$,
where $\sigma_n$ denotes the $n$-th order term in 
an expansion in the gravitational perturbation $h_{\mu\nu}$.
Then, as shown in~\cite{ford}, 
the root-mean-square (RMS) deviation from the classical propagation 
time $\Delta t$ is related gauge-invariantly~\cite{ford} to $<\sigma^2>$ 
by
\be
   \Delta t = \frac{\sqrt{<\sigma^2> -
<\sigma_0^2>}}{L} \simeq \frac{\sqrt{<\sigma_1^2>}}{L} + \dots 
\label{deltat}
\ee
where $L = |x' - x|$ is the distance between the source and the detector.

As commented earlier, one may expect Lorentz invariance to be
broken in a generic theory of quantum gravity, and specifically
in the recoil context discussed in the previous Section. In the
context of string theory, violations of Lorentz invariance entail the
exploration of
non-critical string backgrounds, since Lorentz invariance is 
related to the conformal symmetry that is a
property of critical strings. As we discussed in the previous
Section, a general approach to the formulation of
non-critical string theory involves introducing a 
Liouville field~\cite{distler89} as a conformal factor
on the string world sheet, which has non-trivial dynamics and
compensates the non-conformal behaviour of the string background,
and we showed in the specific
case of $D$ branes that their recoil after
interaction with a closed-string state
produces a local distortion of the surrounding 
space-time (\ref{metric}). 

Viewed as a perturbation about a flat target space-time, 
the metric (\ref{metric}) implies that 
that the only non-zero components of $h_{\mu\nu}$
are:
\be
h_{0i} = \epsilon ^2 {\overline u}_i t \Theta _\epsilon (t)
\label{pert2}
\ee
in the case of $D$-brane recoil.
We now
consider light propagation along the $x$ direction in
the presence of a metric fluctuation $h_{0x}$ (\ref{pert2})
in flat space, along a null geodesic
given by $(dt)^2 = (dx)^2 + 2 h_{0x} dt dx $.
For large times $t \sim {\rm Log}\Lambda/a
\sim \epsilon^{-2}$~\cite{kanti98}, $h_{ox} \sim
{\overline u}$, and thus we obtain
\be
\frac{cdt}{dx}={\overline u} + \sqrt{1 + {\overline u}^2} \sim 
1 + {\overline u} + {\cal O}\left({\overline u}^2\right)
\label{refr}
\ee
where the recoil velocity $\overline u$ is in the direction of the
incoming
light ray. Taking into account energy-momentum 
conservation  in the recoil process, 
which has been derived in this formalism as mentioned previously, 
one has a typical order of magnitude 
${\overline u}/c ={\cal O}(E/M_Dc^2)$,
where $M_D =g_s^{-1}M_s$ is the $D$-brane mass scale, with $M_s \equiv 
\ell_s^{-1}$. Hence (\ref{refr}) implies 
a subluminal energy-dependent velocity of light:
\be
c(E)/c=1 -{\cal O}\left(E/M_Dc^2\right)
\label{vellight}
\ee
which corresponds to a {\it classical} refractive index.
This appears because the metric perturbation (\ref{pert2}) 
is energy-dependent,
through its dependence on ${\overline u}$.

The subluminal velocity (\ref{vellight})
induces a delay in the arrival of a photon of energy $E$
propagating over a distance $L$
of order: 
\be
      (\Delta t)_r =  \frac{L}{c}{\cal O}\left(\frac{E}{M_Dc^2}\right)
\label{figmerclass}
\ee
This effect can be understood physically from the fact  
that the curvature of space-time induced by the recoil is
${\overline u}-$ and hence energy-dependent.
This affects the paths of photons
in such a way that more energetic photons see more
curvature, and thus are delayed with respect to low-energy ones.

As recalled above, the absence of superluminal light propagation
was found previously via the formalism of the Born-Infeld 
lagrangian dynamics of $D$ branes~\cite{mavro+szabo,ellis99}.
Furthermore, the result (\ref{figmerclass}) is in agreement 
with the analysis of~\cite{aemn,amelino98}, 
which was based on a more abstract analysis of Liouville strings.
It is encouraging that this result appears also in this more conventional 
general-relativity approach~\cite{ford}, in which the underlying
physics is quite transparent.

\subsection{Quantum Corrections and Stochastic Effects}

As a preliminary to evaluating
{quantum effects}~\cite{ellis99}
in the context of our simplified model,
we now express $\sigma_1^2$ in terms
of the two-point function of $h_{0x}$,
considering again the null geodesic in the presence of the small metric 
perturbations (\ref{pert2}). To first order in the recoil velocity,
one has: 
$c(\Delta t) = \left(\Delta x  - 2\int _{x_1}^{x_2} dx 
h_{0x}\right)$,  
from which we find
$2\sigma \simeq  (\Delta t)^2 - 
(\Delta x)^2 + 2\epsilon^2 {\overline u}t 
\Theta _\epsilon (t) (\Delta x)^2 $, implying that
$\sigma _1 = \epsilon ^2 {\overline u} t \Theta _\epsilon (t)$.   
One then has:
\be
<\sigma _1^2 > \sim  L^2 \int _x^{x'} \int _x^{x'}
dy \int dy' <h_{0x}(y,t)h_{0x}(y',t')>
\label{metricb}
\ee
In the case of $D$-brane recoil, the computation 
of the quantum average $< \dots >$ may be made in
the Liouville-string approach
described above. In this case, the quantum average 
$< \dots >$ is replaced by a world-sheet correlator calculated
with a world-sheet action deformed by (\ref{logpair}).  
It is clear from (\ref{pert2}) that 
the two-point metric correlator appearing in (\ref{metricb})
is just the $<\De\De>$ world-sheet recoil 
two-point function described in (\ref{twopoint}). 
Thus, at the classical tree level on the world sheet,
one recovers the refractive index
(\ref{figmerclass}) from (\ref{metricb}) by concentrating
on the leading divergence in the $\De \De$ correlator
(\ref{twopoint}), proportional to $\epsilon ^{-2}$. 

To describe fully the quantum effects, one must
sum over world-sheet genera. As discussed in~\cite{emndbrane,mavro+szabo} 
such a procedure results in a canonical quantization 
of the world-sheet couplings, which, in the problem at hand,
coincide with the target-space collective coordinates and momenta
of the recoiling $D$ brane. Thus, the quantum effects arising from 
summation over world-sheet genera result, in a first approximation, in
$<\sigma_1^2 > \sim L^4 \left( \Delta {\overline u}\right)^2 $,
where $\Delta {\overline u}$ denotes the quantum uncertainty 
of the recoil velocity ${\overline u}$, which has been computed 
in~\cite{mavro+szabo}.  

The result of the summation over genera has been discussed
in detail in~\cite{mavro+szabo}, and is not repeated here: we only state
the final results relevant for our purposes. 
The leading contributions 
to the quantum fluctuations in the
space ${\cal M}$ of world-sheet couplings
arise from pinched annulus diagrams 
in the summation over world-sheet genera~\cite{emndbrane,mavro+szabo}. 
These consist of thin tubes of width $\delta\rightarrow 0$, which
one may regard as wormholes attached to 
the world-sheet surface $\Sigma$. The
attachment of each tube corresponds to the insertion of a bilocal pair
of recoil vertex operators 
$V^i(s)V^j(s')$  on the boundary
$\partial\Sigma$, with interaction strength
$g_s^2$, and computing the string propagator along the thin tubes. 
One effect of the dilute gas of world-sheet wormholes
is to exponentiate the bilocal operator, leading to a
change in the world-sheet action~\cite{emndbrane,mavro+szabo}. This
contribution can be cast into the form of a local action by rewriting it as a
Gaussian functional 
integral over wormhole parameters $\rho_i^{ab}$, leading
finally to~\cite{mavro+szabo}:
\be
\sum_{\rm genera}Z \simeq 
\int_{\cal
M}D\rho~e^{-\rho_i^{ab}G^{ij}\rho_j^{cd}/2|\epsilon|^2\ell_s^2
g_s^2\log\delta}~\left\langle W[Y+\rho]\right\rangle_0
\label{genusexp}\ee
where $\langle W[Y+\rho]\rangle_0$
denotes the partition function on a world sheet with the topology of a
disc of a model deformed by the operators (\ref{logpair}),
with couplings shifted by $\rho$, and
$G^{ij}$ is the inverse of the 
Zamolodchikov metric~\cite{zam}
$<~V_i~V_j~>$ in string theory space, evaluated 
on the disc world sheet.
The shifts in the effective couplings $Y_i,u_i$  
imply that they fluctuate statistically,
in much the same way as wormholes and other topology changes in target
space-times
lead to the quantization of the couplings in conventional field 
theories~\cite{coleman}. 

In our case, we see from (\ref{genusexp}) that the effect of this
resummation over pinched
genera is to induce quantum fluctuations in the solitonic $D$-brane
background, giving rise to a
Gaussian statistical spread in the 
collective coordinates of the $D$ brane,  
determined by $G^{ij}$ 
for the logarithmic deformations (\ref{logpair}).
Note that, in such a formalism,
one defines a renormalized string coupling ${\bar g}_s = g_s \epsilon
^{-1}$,
which plays the r\^ole of the physical string coupling in the 
problem~\cite{mavro+szabo}.   
To lowest non-trivial order in the string coupling ${\bar g}_s$
and ${\overline u}^2$, the 
world-sheet analysis
of~\cite{mavro+szabo}, using the methods of logarithmic 
conformal field theory~\cite{lcft}, showed that the quantum fluctuations 
in ${\overline u}$ induced by the summation
over world-sheet topologies are given by;
\be
\left( \Delta {\overline u} \right)^2  = 4 {\bar g}_s^2
\left[1 - \frac{285}{2}{\overline u}^2\right],
\label{delU}
\ee
if the $D$-brane foam corresponds to a 
minimum-uncertainty wave-packet~\cite{mavro+szabo}, 
with the position fluctuations of the $D$ branes
being saturated: $\Delta Y_i \sim {\bar g_s}^{1/3}\ell_s$.  
The energy-independent first part of (\ref{delU})
can be absorbed into $\sigma_0$, and hence
does not contribute to the stochastic 
fluctuations (\ref{deltat}) in the photon arrival time.
The second part of the 
uncertainty (\ref{delU}), which depends on $\overline u$,
and hence the energy of the photon,
cannot be scaled away for all photons of different energies 
by a simple coordinate transformation. 
However, it is of higher 
(second) order in the small parameter ${\overline u}$.
Thus the geodesic correction $\sigma_1$, which is linear in the
gravitational
perturbation $h_{\mu\nu}$, contributes in leading order only to the
classical 
refractive index discussed earlier.

We now repeat the analysis for the quadratic correction $\sigma_2$
in (\ref{deltat}). This
contains the normal-ordered coincidence limit 
of the two-point function of the $D_\epsilon$ operator, which
has a leading-order term
proportional to $\epsilon ^{-2}$ times the $c$-number identity,
as seen in (\ref{twopoint}). Since we work in 
a subtraction scheme in which
one-point functions of the deformations 
(\ref{logpair}) vanish, this means that
$<\sigma_1 \sigma_2> =0$, so that the leading 
contribution comes from the sum over genera of $<\sigma_2^2>$, which
yields fluctuations in the arrival time that are
proportional to the uncertainty $\Delta \left({\overline u}^2\right)
\sim {\overline u}\Delta {\overline u}$.
For a minimum-uncertainty 
state of $D$ branes, we therefore find a contribution 
\be 
| \left( \Delta t\right)_{obs}| \simeq 
{\cal O}\left({\bar g}_s\frac{E}{M_D c^2}\right) \frac{L}{c} 
\label{figuremerit}
\ee
to the RMS fluctuation in arrival times.
As expected, the {\it quantum} effect (\ref{figuremerit}) 
is suppressed by a power of the string coupling constant, 
when compared with the 
classical refractive index effect (\ref{figmerclass}).
The result (\ref{figuremerit}) 
was also derived in~\cite{ellis99} using 
the techniques of Liouville string theory,
via the Born-Infeld lagrangian for the 
propagation of photons in the $D$-brane foam. 
Its derivation here using more
conventional field-theoretic techniques adds robustness to the 
Liouville string theory perspective. It should be
noted that the recoil-induced effect (\ref{figuremerit}) is
larger than the  effects discussed in~\cite{ford},
which are related to
metric perturbations associated with the squeezed coherent states
relevant to particle creation in conventional local field theories.

\subsection{Modification of Maxwell's Equations}

We now consider~\cite{efmmn} the effects of the recoil-induced
space time (\ref{metric2}), viewed as a `mean field solution'
of the $D$--brane-inspired quantum-gravity model,
on the propagation of electromagnetic waves. 
Maxwell's equations in the background metric (\ref{metric2}) 
in empty space can be written as~\cite{landau}:
\bea 
&~& \nabla \cdot  B=0, \qquad \nabla \times H - \frac{1}{c} \frac{\partial
}
{\partial t} D = 0, \nn \\
&~& \nabla \cdot D = 0, \qquad \nabla \times  E
+ \frac{1}{c}\frac{\partial }
{\partial t} B = 0, 
\label{threemx}
\eea
where 
\be 
        D = \frac{E}{\sqrt{h}} + H \times {\cal G}, \qquad 
   B = \frac{H}{\sqrt{h}} + {\cal G} \times E 
\label{media}
\ee
Thus, there is a direct analogy 
with Maxwell's equations in a medium with $1/\sqrt{h}$ 
playing the r\^ole of the electric and magnetic permeability. 
In our case~\cite{emndbrane}, $h=1$, so one has the 
same permeability as the classical vacuum. In the case of
the constant metric perturbation (\ref{metric2}),
after some elementary vector algebra and,
appropriate use of the modified Maxwell's equations,
the equations (\ref{threemx}) read: 
\bea
&~& \nabla \cdot E + {\overline u} \cdot {1 \over c}
\frac{\partial }{\partial t} E 
= 0        \nn \\
&~& \nabla \times B - \left(1 - {\overline u}^2\right) 
{1 \over c} \frac{\partial }{\partial t} E 
+ {\overline u} \times {1 \over c} \frac{\partial}{\partial t } B 
+ \left({\overline u} \cdot \nabla \right) E = 0 \nn \\ 
&~& \nabla \cdot B = 0 \nn \\
&~& \nabla \times E + \frac{1}{c}\frac{\partial}{\partial t}B = 0
\label{mxgrav}
\eea
Dropping non-leading terms of order ${\overline u}^2$
from these equations,
one obtains after some straightforward algebra
the following modified wave equations for $E$ and $B$: 
\bea
&~&\frac{1}{c^2}\frac{\partial ^2}{\partial ^2 t} B - 
\nabla ^2 B - 2 \left({\overline u}. \nabla \right) 
{1 \over c} \frac{\partial }{\partial t} B = 0 \nn \\
&~&\frac{1}{c^2}\frac{\partial ^2}{ \partial ^2 t} E - 
\nabla ^2 E - 2 \left({\overline u}. \nabla \right) 
{1 \over c} \frac{\partial}{\partial t} E = 0 
\label{waves}
\eea
If we consider one-dimensional motion along the $x$ direction,
we see that these equations admit wave solutions of the form
\be
E_x = E_z =0, \; E_y (x,t)= E_0 e^{i(kx - \omega t)}, \;\;
B_x = B_y =0, \; B_z (x,t) = B_0 e^{i(kx - \omega t)},
\label{wave}
\ee
with the modified dispersion relation:
\be
     k^2 - \omega ^2 - 2{\overline u} k \omega = 0
\label{dispr}
\ee
Since the sign of ${\overline u}$ is that of the momentum vector 
$k$ along the $x$ direction, the
dispersion relation (\ref{dispr}) corresponds to 
{\it subluminal} propagation with a refractive index: 
\be
c(E) =c \left(1 - {\overline u} \right)
+ {\cal O}\left({\overline u}^2\right)
\label{refr2}
\ee
where we estimate that
\be
{\overline u} = {\cal O}\left(\frac{E}{M_Dc^2}\right)
\label{refrmag}
\ee
with $M_D$ the $D$-particle mass scale. This is
in turn given by $M_D = g_s^{-1} M_s$ in a string model,
where $g_s$ is the string coupling and $M_s$ is the string
scale~\cite{ellis99}. 
The relation ({\ref{refrmag})
between ${\overline u}$ and the photon energy
has been shown~\cite{mavro+szabo} to
follow from a rigorous world-sheet analysis 
of modular divergences in string theory, but the details need not
concern us here. It merely expresses elementary
energy-momentum conservation, as discussed earlier.

The refractive index effect (\ref{refr2}) is a mean-field effect,
which implies a delay in the arrival times of photons, relative
to that of an idealized low-energy photon for which
quantum-gravity effects can be ignored, of order:
\be
    \Delta t \sim \frac{L}{c} |{\overline u}| = 
{\cal O}\left(\frac{EL}{M_Dc^3}\right)
\label{refrmag2}
\ee 
As we have discussed above (see also~\cite{ellis99}),
one would also expect quantum fluctuations about the mean-field
solution (\ref{refrmag2}), corresponding
in field theory to
quantum fluctuations in the light cone that could be induced by
higher-genus
effects in a string approach. Such
effects would result in 
stochastic fluctuations in the velocity of light 
which are of order 
\be
\delta c \sim 8g_s E/M_Dc^2, 
\label{stochc}
\ee
where $g_s$ is the string coupling, which varies between ${\cal O}(1)$
and $ \ll 1$ in different string models. Such
an effect would
motivate the following parametrization of any possible
stochastic spread in photon arrival times:
\be
\left(\delta \Delta t \right) = {L E \over c\Lambda}
\label{stochc2}
\ee
where the string approach suggests that $\Lambda \sim M_Dc^2 / 8
g_s$.
We emphasize that,
in contrast to the variation (\ref{refr2})
in the refractive index -
which refers to photons of different energy -
the fluctuation (\ref{stochc2}) 
characterizes the statistical spread in the velocities 
of photons {\it of the same energy}. We recall
that the stochastic effect (\ref{stochc2}) 
is suppressed, as compared to the refractive index 
mean field effect (\ref{refrmag}), by an extra power of $g_s$.

\subsection{Consequences for the Propagation of Photon Pulses}

We now discuss~\cite{efmmn} the
propagation of a pulse of photons through
space-time foam (\ref{metric2}). Several
different types of astrophysical sources could be
considered: for orientation, we will bear in mind
GRBs, which typically emit photons in pulses
with a combination of different wavelengths. Their sources
are believed to be ultrarelativistic shocks with Lorentz
factors $\gamma = {\cal O} (100)$~\cite{Piran,Rees}. We do not
enter here into the details of the astrophysical modelling
of such sources, which are unnecessary for our present
exploratory study,
though they may be essential for future more detailed probes of
the constancy of the velocity of light. Instead, here we
study a simple generalization of
the previous discussion of monochromatic wave propagation,
considering a wave packet of photons emitted with a
Gaussian distribution in the light-cone variable $x - ct$.
Since the distance over which the ultrarelativistic source
moves during the emission is negligible compared with the
distance between the source and the observer, we may
represent the source equally well with a Gaussian distribution in $x$ at
the time $t=0$. This is adequate
to see how such a pulse would be modified at the observation point
at a subsequent time $t$, because of the propagation through the 
space-time foam, as a result of the refractive index effect 
(\ref{dispr}), (\ref{refr2}). The phenomenon is similar to the motion 
of a wave packet in a conventional dispersive medium, as discussed 
extensively in the standard literature.

The Gaussian wavepacket may be expressed at $t = 0$ as the real part
of 
\be
f(x) = A e^{-x^2/(\Delta x_0)^2} e^{i k_0 x}
\label{gauss}
\ee
with a modulation envelope that is symmetrical about the origin, where it
has 
amplitude $A$. The quantity $\Delta x_0$ in (\ref{gauss}) denotes the 
root mean square of the spatial
spread of the energy distribution in the packet, which
is proportional to $|f(x)|^2$, as is well known.
If we assume a generic dispersion relation $\omega = \omega (k)$,
a standard analysis using Fourier transforms shows that 
at time $t$ the Gaussian wave-packet will have the form:
\be
   |f(x,t)|^2 = \frac{A^2}{(1 + \frac{\alpha ^2 t^2}{(\Delta x_0)^4})^{1/2}}  
{\rm exp}\{-\frac{(x - c_g t)^2 }{\left(2(\Delta x_0)^2
[1 + \frac{\alpha ^2 t^2}{(\Delta x_0)^4}]\right)}\}
\label{timegauss}
\ee
where $\alpha \equiv \frac{1}{2} \left(d^2\omega /d^2 k\right)$,
and $c_g \equiv d\omega/dk$ is the group velocity, i.e.,
the velocity with which the peak of the distribution moves
in time. 

We see immediately in (\ref{timegauss}) that the quadratic term 
$\alpha$ in the dispersion relation
does not affect the motion of the peak, but only the spread of the
Gaussian wave packet:
\be
 | \Delta x | = 
\Delta x_0 \left( 1 + \frac{\alpha ^2 t^2}{(\Delta x_0)^4}\right)^{1/2}
\label{spread}
\ee
which thus increases with time. The quadratic term $\alpha$ also affects 
the amplitude of the wave packet:
the latter decreases together with the increase in the spread
(\ref{spread}),
in such a way that the integral of $|f(x,t)|^2$ is constant. 

In the case of the quantum-gravitational foam scenario 
of~\cite{ellis99,aemn}, the dispersion relation 
assumes the following form for positive momentum $k$:
\bea
&~& k=\omega \left( 1 + \frac{\omega}{M_D}\right) 
~{\rm or} ~ \omega = k \left( 1 - \frac{k}{M_D} + \dots \right) \nn \\
&~&c_g = \left(1 - {\overline u}\right) = 1 -
{\cal O}\left(\omega/M_D\right), \nn \\
&~&\qquad \alpha = - \frac{1}{M_D} + \dots 
\label{quant}
\eea
where we denote by $\dots$ the higher-order (e.g., quadratic)
terms in $1/M_D$, which are subleading in this case.   
Thus the spread of the wave packet due to the non-trivial  
refractive-index effect described earlier is:
\be
    | \Delta x| = \Delta x_0 
\left( 1 + \frac{t^2}{M_D^2 (\Delta x_0)^4} \right)^{1/2} 
\label{light}
\ee
We note that the spread due to the refractive index $\delta c/c \propto
\omega$ is independent of the energy of
the photon to leading order in $1/M_D$. 
We also note, therefore, that this effect is distinct from the
stochastic propagation effect, which gives rise to a spread 
(\ref{stochc2}) in the
wave-packet that depends on the photon energy $\omega$.
For astrophysical sources at cosmological distances
with redshifts $z \simeq 1$, and with an initial $\Delta x_0
$ of a few km, one finds that the correction (\ref{light})
is negligible if the quantum-gravity 
scale $M_D$ is of the order of $10^{19}$ GeV, namely of order 
$10^{-30}\Delta x_0$.
The correction would become of order $\Delta x_0$ only if the
latter is of order $10^{-3}$~m. Even if one allows $M_D$ to be as
low as the sensitivities shown in Table 1, this broadening effect
is still negligible for all the sources there, being at most of
order $10^{-22} \Delta x_0$. Therefore, in this particular
model, the only broadening effect that needs to be considered
is the stochastic quantum-gravitational effect on the 
refractive index that was introduced at the end of the previous
subsection.

{\small 
\begin{table}[ht]
\begin{center}
\begin{minipage}{\linewidth}
\renewcommand{\thefootnote}{\fnsymbol{footnote}}
\begin{center}
{\bf Table~1:~Observational~Sensitivities~and~Limits~on~$M$}
\end{center}
\begin{tabular}{|c||c|c|c|c|}   \hline
Source & Distance & $E$ & $\Delta t$ & 
\parbox{3cm}{Sensitivity to $M$} 
\\ \hline
GRB 920229~{\cite{amelino98,schaefer}} & 3000
Mpc (?) & 200 keV & $10^{-2}$
s &
$0.6\times 10^{16}$ GeV (?)
\\ \hline
GRB 980425~$^a$ & 40 Mpc & 1.8 MeV & $10^{-3}$ s (?)
& 
$0.7\times 10^{16}$ GeV (?)
\\ \hline
GRB 920925c~$^a$ & 40 Mpc (?) & 200 TeV (?) & 200 s & 
$0.4\times 10^{19}$ GeV (?)
\\ \hline
Mrk 421~{\cite{biller}} & 100 Mpc & 2 TeV & 280 s & 
$> 7 \times 10^{16}$ GeV
\\ \hline
Crab pulsar~{\cite{crab}} & 2.2 kpc & 2 GeV & 0.35 ms & $> 1.3 \times
10^{15}$ GeV
\\ \hline
GRB 990123 & 5000 Mpc & 4 MeV & 1 s (?) & $2 \times 10^{15}$ GeV (?)
\\ \hline
\end{tabular} 
\end{minipage} 
\end{center}  
\caption{{\it The
mass-scale parameter $M$ is defined by $\delta c/c = E/M$.
The question marks in the Table indicate uncertain 
observational inputs. Hard limits are indicated by inequality signs.}}
\label{tabl1} 
\end{table}
}
\vspace{0.5cm}

In the case of a different quantum-gravitational foam scenario
with a quadratic refractive index: $\delta c / c \sim E^2$,
the dispersion relation would take the following form:
\bea
&~& k=\omega \left( 1 + \frac{\omega}{\tilde{M}}\right) 
~{\rm or} ~ \omega = k \left( 1 - \left(\frac{k}{\tilde M}\right)^2 + 
\dots \right) \nn \\
&~&c_g = = \left(1 - {\overline u}\right) = 
1 - {\cal O}\left(\omega^2/{\tilde M}^2\right), \nn \\
&~&\qquad \alpha = - 3\frac{\omega}{{\tilde M}} + \dots 
\label{quant2}
\eea
where $\dots$ again denote subleading terms.   
In this case, 
the spread of the wave packet due to the non-trivial  
refractive index effect described above is:
\be
    | \Delta x| = \Delta x_0 
\left( 1 + \frac{9 \omega^2 t^2}{{\tilde M}^4 (\Delta x_0)^4} \right)^{1/2} 
\label{light2}
\ee
Once again, if one takes into account the sensitivities shown
in Table 1, the maximum spreading of the pulse is negligible
for $\Delta x_0 \sim 10^{-3}$~m, namely at most $\sim 10^{-33} \Delta
x_0$. Once again, one would need only to consider the possible
stochastic quantum-gravitational effect on the refractive
index. However, since a quadratic dependence is
not favoured in our particular theoretical approach, we do not pursue it
further in the rest of this paper.

\subsection{Cosmological Expansion} 

We now discuss the implications of the cosmological expansion for the
searches for a quantum-gravity induced refractive index (\ref{refrmag2}) 
and the stochastic effect (\ref{stochc2}).
We work within the general context
of Friedmann-Robertson-Walker (FRW) metrics, 
as appropriate for standard homogeneous and
isotropic cosmology~\cite{wein}.
We denote by $R$ the FRW scale factor,
adding a subscript $0$ to denote the value at the present era,
$H_0$ is the present Hubble expansion parameter,
and the deceleration parameter
$q_0$ is defined 
in terms of the curvature $k$ of the FRW metric by
$k = (2q_0 - 1) (H_0^2 R_0^2 /c^2)$,
i.e., $\Omega _0 =2q_0$. 

Motivated by inflation and the cosmic microwave background data,
we assume a Universe with
a critical density: $\Omega _0=1$, $k=0$ and $q_0=1/2$.
We also assume that the Universe is matter-dominated
during all the epoch of interest.
Then the scale factor $R(t)$ of the Universe expands as:
\be
{R(t) \over R_0} = \left({3H_0 \over 2}\right)^{2/3} t^{2/3} 
\label{exp}
\ee
and the current age of the Universe is 
\be
 t_0 =\frac{2}{3 H_0}
\label{age}
\ee
Clearly no time delay can be larger than this.
The relation between redshift and scale factor is
\be
R(t)/R_0 =1/(1 + z) 
\label{redshift}
\ee
Substituting (\ref{redshift}) into (\ref{exp}),
we find the age of the Universe at any given redshift:
\be
   t=(\frac{2}{3H_0})\frac{1}{(1+ z)^{3/2}}=\frac{t_0}{(1+ z)^{3/2}}
\label{univage}
\ee
Hence, a photon (or other particle) emitted 
by an object at redshift $z$ has travelled for a time 
\be
 t_0 - t =\frac{2}{3H_0} \left(1 -\frac{1}{(1+z)^{3/2}}\right)
\label{tizero}
\ee
The corresponding differential relation between time and 
redshift is
\be
    dt = -\frac{1}{H_0}\frac{1}{(1 + z)^{5/2}}dz 
\label{diff}
\ee
This means that 
during the corresponding infinitesimal time (redshift) interval,
a particle with velocity $u$ travels a distance
\be
u\,dt = -\frac{1}{H_0}\frac{u}{(1 + z)^{5/2}}dz.
\ee
Therefore, the total distance $L$ travelled by such a
particle since emission at redshift $z$ is
\be 
     L = \int_t^{t_0} u dt = 
\frac{1}{H_0}\int _0^z \frac{u(z)}{(1 + z)^{5/2}}dz. 
\label{distance}
\ee
Hence the difference in distances covered by two particles
with velocities differing by $\Delta u$ is: 
\be
   \Delta L =\frac{1}{H_0} \int _0^z \frac{dz}{(1+ z)^{5/2}}(\Delta u)
\label{du}
\ee
where we allow $\Delta u$ to depend on $z$.

In the context of our quantum-gravity-induced refractive-index phenomenon
(\ref{refr2}), we are confronted with just such a situation.
Consider in that context two photons travelling with
velocities very close to $c$,
whose present-day energies are $E_1$ and $E_2$. 
At earlier epochs, their energies would have been blueshifted
by a common factor $1 + z$. Defining $\Delta E_0 \equiv E_1 - E_2$,
we infer from (\ref{refrmag}) that $\Delta u = (\Delta E_0 \cdot (1 +
z))/M$.
Inserting this into (\ref{du}), 
we find an induced difference
in the arrival time of the two photons given by
\be
  \Delta t =\frac{\Delta L}{c} \simeq \frac{2}{H_0}\left[ 1 - \frac{1}{(1 + z)^{1/2}}\right] 
\frac{\Delta E_0}{M}. 
\label{deltacosm}
\ee
The expression (\ref{deltacosm}) describes the corrections 
to the refractive index effect (\ref{refrmag2})
due to the cosmological expansion. 
For small $z << 1$, the general expression (\ref{deltacosm}) 
yields $\Delta t \simeq (z \cdot \Delta E_0)/ (H_0 \cdot M)$,
which agrees with the simple expectation
$\Delta t \simeq (r \cdot \Delta E_0) / (c \cdot M)$
for a nearby source at distance $r = c(t_0 - t) \simeq
{cz}/{H_0} + \dots $.  
There would be similar cosmological corrections to the 
stochastic effect (\ref{stochc2}), also
given by an expression of the form (\ref{deltacosm}),
but with $\Delta E_0 \rightarrow E$, $M \rightarrow \Lambda$,
where $E$ is a typical energy scale in a single channel.

\subsection{Sample Analysis of Data from GRBs}

\begin{figure}[htb]
\epsfxsize=3in
\bigskip
\centerline{\epsffile{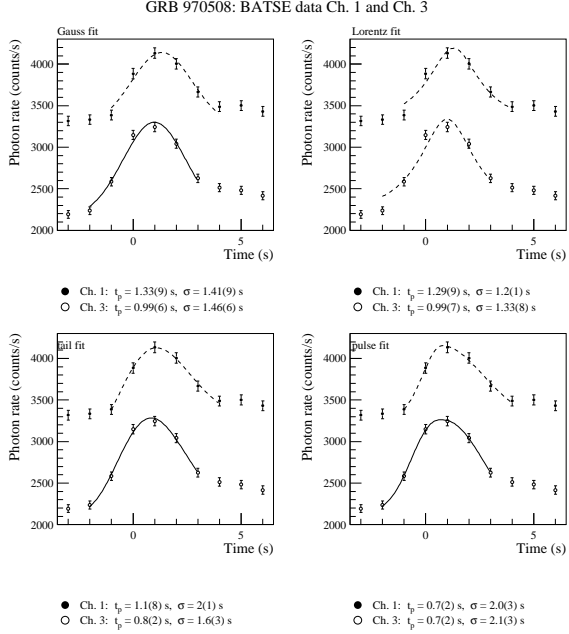}}
\caption{\it\baselineskip=12pt Time distribution 
of the number of photons 
observed by BATSE in Channels 1 and 3 for GRB~970508, compared 
with the following fitting functions~\cite{efmmn}: (a) Gaussian, (b)
Lorentzian, (c) `tail' function, and (d) `pulse' function.
We list below each panel the positions $t_p$ and widths $\sigma_p$
(with statistical errors) found for each peak in each fit. We
recall that the BATSE data are binned in periods of 1.024~s.}
\label{figdat1}\end{figure}

We presented in~\cite{efmmn} a detailed 
analysis of the astrophysical data for 
a sample of 
Gamma Ray Bursters (GRB) 
whose redshifts $z$ are known (see Fig. \ref{figdat1}
for the data of a typical burst: GRB~970508).
We looked (without success) 
for a correlation with the redshift,
calculating a regression measure (see Fig. \ref{figregrz}) for the 
effect (\ref{deltacosm}) and
its stochastic counterpart (\ref{stochc2}). 
Specifically, we looked for linear dependences of
the `observed' $\Delta t/\Delta E_0$ and the spread 
$\Delta \sigma/E$ on ${\tilde z}
\equiv 2 \cdot [ 1 - (1 / (1 + z)^{1/2}] \simeq z - (3/4) z^2 +
\dots$. 
We determined limits on
the quantum gravity scales $M$ and $\Lambda$ 
by constraining the possible magnitudes of the slopes  in 
linear-regression analyses of the differences between the arrival times
and widths of pulses in different energy ranges from five GRBs with
measured redshifts,
as functions of ${\tilde z}$ (\ref{deltacosm}).
Using the current value for the Hubble expansion parameter,
$H_0 = 100 \cdot h_0$~km/s/Mpc, where $0.6 < h_0 < 0.8$,
we obtained 
the following limits~\cite{efmmn}
\be 
M \gsim 10 ^{15} \; {\rm GeV}, \quad \Lambda \gsim 2 \times 10 ^{15} \;
{\rm GeV} 
\label{limit}
\ee
on the possible quantum-gravity effects.

\begin{figure}[htb]
\epsfxsize=3in
\bigskip
\centerline{\epsffile{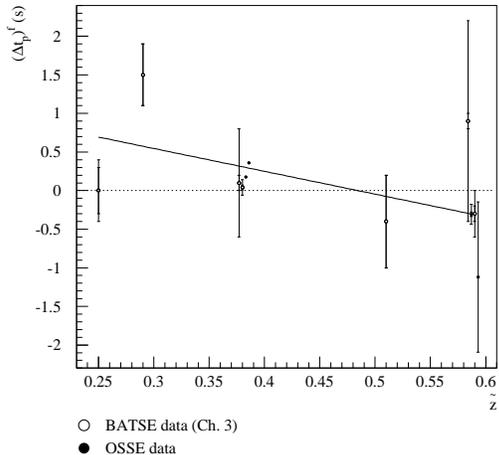}}
\caption{\it\baselineskip=12pt Values of the shifts $(\Delta t_p)^f$ in
the timings of the peaks fitted for each GRB studied
using BATSE and OSSE data~\cite{efmmn}, plotted versus 
${\tilde z}=1-(1+z)^{-1/2}$, where $z$ is the 
redshift. The indicated errors are the
statistical errors in the `pulse' fits provided by the fitting routine,
combined with systematic error estimates obtained
by comparing the results obtained using the `tail' fitting
function. The values obtained by comparing OSSE with BATSE 
Channel 3 data
have been rescaled by the factor $(E_{min}^{BATSE~Ch.~3} -
E_{max}^{BATSE~Ch.~1}) / (E_{min}^{OSSE} - E_{max}^{BATSE~Ch.~3})$,
so as to make them directly comparable with the comparisons of
BATSE Channels 1 and 3. The solid line is the best linear fit.}
\bigskip
\label{figregrz}\end{figure}

\section{Other Phenomenological Issues}

\subsection{Extension to Massive Relativistic Particles} 

So far, we have concentrated our attention on massless
particles, namely photons, because of the very interesting
experimental/observational possibilities they provide. 
It should however be clear, that the metric
perturbation (\ref{metric2}) produced by our $D$-brane recoil model,
which implies
a breakdown of Lorentz invariance, alters the Einstein dispersion 
relation for massive particles too. One expects, on general grounds,
that the photon dispersion relation,
\begin{equation}
\omega^2 - k^2  + 2 \bar{u}k\omega = 0
\label{pdr}
\end{equation}
will become in the case of massive particles,
\begin{equation}
\omega^2 - k^2  + 2 \bar{u}k\omega - m^2_o = 0\, ,
\label{mdr}
\end{equation}
leading to the modified Einstein realtion,
\begin{equation}
\omega^2\approx (k^2 + m^2_o)\left[ 1 - \frac{k^2}{\sqrt{k^2 + m^2_o}}
                                        \frac{1}{M}\right]^2
\label{mera}
\end{equation}
to leading order in ${\cal O}\left(\frac{k}{M}\right)$.
It is useful to recast (\ref{mera}) in the more familiar
notation: 
\begin{equation}
E^2= (p^2 + m^2_o)\left[ 1 - 
     \left(\frac{1}{\sqrt{1 + \frac{m^2_o}{p^2}}}\right)
                                        \frac{p}{M}\right]^2\, ,
\label{merb}
\end{equation}
which in the non-relativistic limit yields:
\begin{equation}
E = m_o + \frac{p^2}{2m_o} -  \frac{p^2}{M} 
        + \cdots {\cal O}\left( \left(\frac{p}{M}\right)^2 \right),
\label{nrl}  
\end{equation}
Thus, an effective rest mass
\begin{equation}
(m_{eff})_o \approx m_o \left( 1 + \frac{2 m_o}{M} \right)\, .
\label{erm}
\end{equation}
appears in the non-relativistic kinetic energy:
$E_{kin} = \frac{p^2}{2m_{eff}}$. Alternatively,
one may use (\ref{merb}) to recast the T-shirt
formula $E= mc^2$ in the form
\begin{equation}
E\approx mc^2 \left[ 1 - \frac{m}{M}\left(\frac{{\overline u}^2}{c^2}\right)\right],
\label{tte}
\end{equation}
where $m\equiv \frac{m_o}{\sqrt{1-{\overline u}^2/c^2}}$, as usual.

We see again that although quantum gravitational fluctuations in 
spacetime, as modelled by $D$-brane quantum recoil, may lead to a
spontaneous
breakdown of Lorentz symmetry, the resulting corrections
to the standard Einstein relations are very small, so that
conventional Special Relativity is still a good 
approximation to the world.

Before closing this subsection, we recall that another
possible probe of quantum-gravitational effects on massive particles is
offered by tests of quantum mechanics in the neutral kaon system. 
A parametrization of possible deviations
from the Schr\"odinger equation has been given~\cite{ellis84},
assuming energy and probability conservation, in terms of
quantities $\alpha,\beta,\gamma$ that
must obey the conditions
\begin{equation}
\alpha , \, \gamma \, > \, 0, \qquad \alpha \gamma \, > \, \beta^2
\label{positiv}
\end{equation}
stemming from the positivity of the density matrix $\rho$. 
These parameters 
induce quantum decoherence and violate CPT~\cite{ELMN}. 
Experimental data on neutral kaon decays so far agree perfectly with
conventional
quantum mechanics, imposing only the following upper limits~\cite{emncplear}:
\begin{equation}
\alpha < 4.0 \times 10^{-17} \hbox{GeV}, \qquad 
\beta < 2.3 \times 10^{-19} \hbox{GeV}, \qquad 
\gamma < 3.7 \times 10^{-21} \hbox{GeV}
\label{bounds}
\end{equation}
We cannot help being impressed that these bounds are in the
ballpark of $m_K^2 / M_P$, which is the maximum 
magnitude that we could expect any such effect to have.

This and the example of photon propagation give hope that 
experiments may be able to probe physics close to the Planck scale,
if its effects are suppressed by only one power of $M_P \simeq
10^{19}$~GeV. One
should not exclude the possibility of being able to
test some of the speculative ideas about quantum gravity reviewed
in this article. Indeed, {\it if} the analysis of photon
propagation can be extended to energetic neutrinos, and {\it if}
GRBs emit $\sim 10$~s $\nu$ pulses, one could be sensitive to
mass scales as large as $10^{28}$~GeV!

\subsection{$D$--particle Recoil and Time-Dependent Vacuum Energy} 

We would now like to describe briefly another interesting 
consequence~\cite{cosmol} of our recoil formalism, namely
the generation of a time-dependent vacuum energy.
We recall that the possibility of non-zero vacuum energy
has recently received dramatic support from an
unexpected source, namely observations of high-redshift supernovae~\cite{sn}.
These indicate that the large-scale geometry of the Universe is not
that of a critical matter-dominated cosmology, and that its
expansion may even be accelerating. The supernova data are consistent
with $\Omega_{\Lambda} \sim 0.7$, if the Universe is indeed close to
critical as suggested by inflation. The supernova and other data
are beginning to provide interesting constraints on the
equation of state of this vacuum energy, which is consistent with
being a cosmological constant, but some time dependence in the
vacuum energy is also consistent with the data.

For the purposes of this subsection,
we consider cosmology in the framework of a `material reference
system' (MRS) of $D$ particles~\cite{cosmol},  
including the effects of quantum recoil
induced by closed-string probe particles. 
For reasons of supersymmetry, we expect that the 
the ground state of the quantum-mechanical system of closed strings
and $D$ particles should have zero energy,
if recoil fluctuations of the $D$ particles are ignored,
in which case there would be no cosmological vacuum energy.
However, the findings of~\cite{cosmol}, which we recapitulate briefly
below, 
indicate that, due to the recoil process,
the vacuum energy of the quantum-mechanical 
string/$D$--particle 
system is non-zero at finite times, and only
relaxes to zero asymptotically in target time.
As discussed in detail in~\cite{emndbrane}, 
the recoil fluctuations are world-sheet higher-genus effects, and as such 
may be viewed as quantum fluctuations about a supersymmetric 
ground state of ${\cal D}$ particles and strings. The emergence of 
a cosmological constant may be interepreted
as a supersymmetry-breaking contribution of the $D$--particle 
recoil,
which however relaxes to zero asymptotically in (cosmic) time.  
Notice that the entire recoil formalism we adopted in \cite{emndbrane}
is a non-equilibrium decoherening process, and from this point of view
the emergence of time-dependent supersymmetry-breaking contributions 
to the 
ground-state energy should not be too shocking. 

As mentioned earlier, the closed-string/$D$-particle system 
can be compared with
that of an ion lattice in a solid (c.f. ${\cal D}$-particles )
with valence electrons (c.f. closed-string states)
moving freely. 
Lattice ion vibrations (c.f. recoil of ${\cal D}$-particles)  
induce effective phonon interactions which bind the 
electrons resulting in BCS superconductivity for the ground state
of this many-body system. 
In this example, the physical excitations above this ground state
are quasiparticles, which are not the ordinary electrons.
In the $D$--brane analogy, the quantum recoil induced by
the scattering of a closed-string state on a defect
{binds} the string to the defect, creating a collective 
recoil excitation of
the system~\cite{emndbrane}.
Such excitations
are related formally to higher
world-sheet topologies of the string, 
i.e., quantum effects in target space-time, which
break target-space supersymmetry and make a non-trivial 
contribution to the ground-state energy, as discussed below. 
However, due to the non-equilibrium nature of the process,
the recoil ground-state contribution relaxes to zero 
asymptotically in target time.

As in previous sections, we 
concentrate on a single scattering event, namely
the scattering of a single closed-string state by a single defect
as the first step
in a dilute-gas approximation for the ${D}$ particles, which should
be sufficient to describe qualitatively the 
leading behaviour of the vacuum energy of the Universe. 
We are unable at present to treat fully the
more realistic
case of an ensemble of defects with Planckian density, 
due to our limited understanding of the underlying 
microscopic dynamics. 

We consider the very low-energy 
limit ${\overline u}_i \rightarrow 0$ in which
the only non-vanishing components of the 
$D+1$-dimensional Ricci tensor, corresponding to the metric
(\ref{metric}), are~\cite{kanti98}: 
\be
R_{ii} 
\simeq 
\frac{-(D-1)/|\epsilon|^4}{(\frac{1}{|\epsilon|^4} - \sum_{k=1}^{D}|y_i|^2)^2} 
+{\cal O}(\epsilon ^{8}) 
\label{limRicci}
\ee
If we view (\ref{limRicci}) as the ${\cal O}(\alpha ')$ (where $\alpha ' $
is the Regge slope) 
world-sheet $\beta$ function of an appropriate 
stringy $\sigma$ model with target-space metric
perturbations, then, in this limiting case, 
the time (time $t$) decouples when $t >>0$, and 
one is left with a spatial 
$D$-dimensional manifold alone. Thanks to the
symmetries of the string construction, this manifold is, moreover,
maximally symmetric. Therefore one may reconstruct the
metric from the Ricci tensor (\ref{limRicci}):
\be 
   R_{ij}=\frac{1}{D}{\cal G}_{ij}R 
\label{newricci}
\ee
where ${\cal G}_{ij}$ is a diagonal metric corresponding to the line
element:
\be
      ds^2=\frac{|\epsilon|^{-8}\sum_{i=1}^{D} dy_i^2}{(\frac{1}{|\epsilon|^4} - \sum_{i=1}^{D}|y_i|^2)^2} 
\label{ball}
\ee
This metric describes the interior
of a $D$-dimensional ball, 
which is the Euclideanized version of an 
anti-de-Sitter (AdS) space time. In its Minkowski version,
one can easily check that the curvature corresponding to (\ref{ball}) is
\be
R = -4 D (D - 1) |\epsilon|^4,
\label{curvature}
\ee
which is {\it constant} and {\it negative}. The radius of the AdS
space is $b = |\epsilon|^{-2}$.

The Ricci tensor (\ref{limRicci})
cannot be a consistent string background
compatible with conformal invariance to order 
$\alpha '$ if only tree-level world-sheet 
topologies are taken into account. 
However, as shown in~\cite{fischler},
this conclusion no longer holds when
one includes string-loop corrections. These induce a target-space
cosmological constant, corresponding to a dilaton 
tadpole, which renders the backgrounds (\ref{limRicci}) consistent 
with the conformal-invariance conditions.

Alternatively, as discussed
in~\cite{kanti98}, the cosmological vacuum energy
may be obtained from an effective
tree-level non-critical Liouville string 
with central-charge deficit 
\be 
     Q^2 = \Lambda \propto -2(\alpha  ')^2 
(D-1)(D-2)|\epsilon |^4 + {\cal O}(\epsilon ^6)
\label{ccd}
\ee
As it stands, the resulting 
non-conformal stringy $\sigma$ model (\ref{ball})
has a central charge deficit (\ref{ccd}) which is negative for the
expected real values of $\epsilon^2$,
and hence corresponds to a {subcritical} string 
theory~\cite{aben}.
The Liouville mode $\varphi$ 
required to dress~\cite{distler89} this deformed $\sigma$ model
should not be confused with the target time 
variable that appears in the previous Section. 
The $\sigma$-model kinetic term for this  
Liouville field $\varphi$ is:
\be
    -Q^2 \int _{\Sigma} d^2\sigma \partial \varphi \partial \varphi 
\label{kinetic}
\ee
with $Q^2 < 0$ given by (\ref{ccd}), from which
we see that $\varphi$ is an extra {\it space-like} target-space
coordinate. 
Therefore, the $D+1$-dimensional string description
resembles a conventional Euclideanized formulation
of quantum gravity.

Thus, we make a conventional Wick rotation to identify the 
{analytic continuation}: $\varphi \rightarrow i\varphi $ 
as a Liouville `physical' time $t$. Equations (\ref{ccd})
and (\ref{kinetic}) then imply that 
$\epsilon$ should itself be analytically continued
in order to obtain a time-like Liouville theory:
\be 
     t \sim i\epsilon^{-2}
\label{physical}
\ee
With this identification, the spatial manifold (\ref{ball}) is
identified as a spatial slice of a space-time with
signature $(D,1)$. In our approach, slices can be ordered in $i \varphi$
just as they would be ordered in $t$ in a canonical approach to
conventional quantum gravity.
Explicit Lorentz covariance is sacrificed in any such splitting.
However, in our case this should be regarded as a
spontaneous breaking, in the sense that 
it is a property of the ground state of the closed-string/$D$-brane 
theory. Since this ground state is spatially homogeneous
and isotropic, it is {\it a priori} suitable for describing a
Friedmann-Robertson-Walker (FRW) type of cosmology. This is
normally thought to be invariant under local Lorentz
transformations, but it also includes the possibility
that there may be a preferred Machian frame~\footnote{Such as that
provided in the present Universe by the microwave background
radiation.}, as is the case in our approach.

The consistency of the 
resulting metric $G_{\mu\nu}^{ph}$, $\mu,\nu=1, \dots D+1$, 
with Einstein's equations has non-trivial consequences. 
As the the analysis~\cite{cosmol} 
indicates, the physical 
Universe is of FRW type with a scale factor 
\be 
    R(t) = t^2 
\label{scalecs}
\ee
This can be contrasted with the tree-level cosmological model
of~\cite{aben}, where a linear expansion was found as
$t \rightarrow \infty$. 
We see from (\ref{ccd}) and (\ref{physical}) 
that the Universe (\ref{scalecs}) has a time-dependent 
vacuum energy $\Lambda(t)$ which relaxes to zero as:
\be 
      \Lambda (t) = \Lambda (0) /t^2 = 1/R(t)
\label{lambdaphys}
\ee
In accordance to the standard Einstein's equation,
this time-varying {positive} 
vacuum energy drives the cosmic expansion:
\be
\left(\frac{{\dot R}(t)}{R(t)} \right)^2 = \frac{1}{3} \Lambda (t) 
\label{einstein}
\ee
where the dot denotes a derivative with respect to the 
physical (Einstein) time $t$. 

We emphasize that the 
cosmological background described above satisfies
Einstein's equations, and therefore can be interpreted as a
consistent $D+1$-dimensional string background. It is 
{\it not} characterized 
by a cosmological {\it constant}, but by a {\it time-dependent} vacuum 
energy. This time dependence 
has an arrow~\cite{emn}, just like conventional FRW cosmology,
which stems in this approach from the irreversibility of the world-sheet
renormalization-group flow in two-dimensional field theories~\cite{zam,emn}.
This background may be
interpreted as breaking
Lorentz covariance spontaneously,
via the non-trivial vacuum properties of the material reference 
frame of the $D$ particles.

It should be stressed that the above effect
is a very low-energy contribution, appearing when the 
recoil velocity 
${\overline u}_i$ is ignored. As we have seen 
in the above construction, the $\sigma$ model
with a target space (\ref{scalecs}) 
{is not} the original $\sigma$ model
for the metric (\ref{metric}), but one 
whose $\beta$ function reproduces, in the 
${\overline u}_i \rightarrow 0$ limit,  
the limiting Ricci tensor (\ref{limRicci}). 
In this respect, the cosmology described above may be related
to the recoil-induced space-time (\ref{metric}) 
by a marginal deformation.
The further effects on this cosmology 
of a non-trivial recoil velocity, as appropriate
for the metric (\ref{metric2}) studied in previous sections, 
remain to be studied.  
 
\subsection{$D$--brane Recoil and `Large' Extra Dimensions}

Before closing, we would like to describe briefly
another possible application of the 
recoil approach to $D$--brane gravity.
This last application is inspired by the 
recent suggestion~\cite{dimopo98} that the observable 
world is actually a $D3$ brane embedded in a higher-dimensional
(bulk) space-time, so that only closed-string states in the
gravitational supermultiplet can propagate in the bulk.
Recoil effects like those described above appear in
this case as well, since the non-rigid character of the 
$D3$--brane shows up when it emits closed string states into
the bulk, as illustrated in Fig.~3. 
This problem has been studied in~\cite{cemn}, to which
we refer the reader for further details. 
For the  present purposes, we shall summarize only the main result
on light-cone broadening due to the 
recoil fluctuations of the $D3$ brane.
The geometry considered is depicted in Fig.~3:
it consists of two parallel $D3$ branes, one of which represents
our world, separated by a distance $\Lambda$.
The other brane could be a parallel world in the 
sense of~\cite{dimopo98}, or it be another copy of our world,
obtained
upon compactifying the bulk dimension, i.e., imposing periodic
boundary conditions for the gravitons. We consider propagation
of photons (massless particles) along the $D3$ brane, but, in order 
to capture the quantum uncertainty in the $D3$ brain position,
we suppose that the photons travel at a distance $\ell_s$ (the string
length) from the brane. 

\begin{figure}
\begin{center}
\begin{picture}(155,155)(0,-5)
\Oval(28,78)(20,10)(0)
\CBox(28,58)(38,98){White}{White}
\Line(28,58)(50,78)
\Line(28,98)(50,78)
\Vertex(49.8,78){.3}
\Line(10,10)(10,110)
\Line(10,10)(45,45)
\Line(10,110)(45,145)
\Line(45,45)(45,73.5)
\Line(45,82.5)(45,145)
\Line(110,10)(110,110)
\Line(110,10)(145,45)
\Line(110,110)(145,145)
\Line(145,45)(145,145)
\LongArrow(60,5)(110,5)
\LongArrow(60,5)(10,5)
\Text(60,12)[c]{$l=\pi\Lambda$}
\Text(12,33)[l]{$D_1$}
\Text(112,33)[l]{$D_2$}
\PhotonArc(75,78)(12,0,360){1.7}{9}
\LongArrow(80,78)(100,78)
\DashLine(80,78)(63,78){2}
\Line(63,78)(51,78)
\DashArrowLine(28,33)(28,60){3}
\DashArrowLine(28,96)(28,123){3}
\DashArrowArcn(74.5,78)(50,200,160){3}
\end{picture}
\caption{\it Schematic representation of the recoil effect
in a model with `large' extra dimensions:  the photon's
  trajectory (dashed line) is distorted by the conical singularity in
  the brane that results from closed--string emission into the
  bulk.\label{fig3}}
\end{center}
\end{figure}
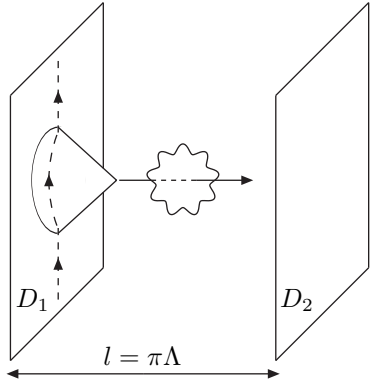

The result for the light-cone fluctuations, as discussed 
previously, is: 
\begin{eqnarray}
  \frac{\Delta t}{t_p} \sim \frac{2^{n/2}}{2\pi}
  \left(\frac{L}{\ell_p}\right)^{1/2} \sqrt{\ln [L/\ell_p]}\;
  \left(\frac{\ell_p}{\pi\Lambda}\right)^{n/4},
\label{twothirteen}
\end{eqnarray}
which for the plausible case $n\!=\!6$ reduces to:
\begin{eqnarray}
  \frac{\Delta t}{t_p} \sim \frac{4}{\pi^{5/2}}
  \left(\frac{L}{\Lambda}\right)^{1/2}
  \left(\frac{\ell_p}{\Lambda}\right) \sqrt{\ln[L/\ell_p]}.
\label{twofourteen}
\end{eqnarray}
The ratio $\ell_p/\pi\Lambda$ can be calculated in terms of the
fundamental Planck scale $M_p^{(4+n)}$, as follows.
Following~\cite{dimopo98}, we have
\begin{eqnarray}
  \ell_p &=& 2\times 10^{-17} \left(\frac{1 {\mathrm{TeV}}}{M_p^{(4+n)}}
  \right) {\mathrm cm}\nonumber\\
  \pi\Lambda &=& 10^{30/n - 17} \left(\frac{1 {\mathrm{TeV}}}{M_p^{(4+n)}}
  \right)^{1+2/n} {\mathrm cm}
\end{eqnarray}
whence, if $M_p^{(4+n)}= \mu \, {\mathrm{TeV}}$,
\begin{eqnarray}
  \left( \frac{\ell_p}{\pi\Lambda}\right) = 2 \times 10^{-30/n} \mu^{2/n}.
\end{eqnarray}
For $n=6$, this ratio is bigger than \(2\times 10^{-5}\) and grows with
$\mu$ as \(\mu^{1/3}\), and for $\mu\sim 100$ it can be as large as
$10^{-4}$.

To gain an understanding whether the light--cone broadening effect
might in principle be measurable, we consider some specific cases.
For a GRB with redshift $z \sim 1$, $L\sim 10^{28}
\mathrm{cm}$, and, using the above formulae, we estimate $\Lambda\sim
10^{-12} \mathrm{cm}$, $\ell_p\sim 10^{-17} \mathrm{cm}$ for $ n = 6$,
and hence
\begin{eqnarray}
  \Delta t \sim 10^{15} t_p,
\end{eqnarray}
In the extra-dimension picture the fundamental Planck time is very
much larger than usually supposed in four dimensions, being of order
$10^{-27}$ seconds, so that
the light--cone broadening is \(10^{-12}\) seconds. 
However, this is still far below
the sensitivity of experiments measuring $\gamma$--ray
bursts~\cite{amelino98}, which is in the millisecond region.  It is
easy to see that the effect is even smaller for $ n < 6$.  The above
estimates have been made for $\mu\!=\!1$: if the underlying scale is
significantly higher than this, both estimates would get larger.  In
the case of gravity--wave interferometers, the sensitivity of the
experiments is much better~\cite{amelino99}, namely of the order of
$10^{-18}$ metres.  For this case we have $L\sim 10^{3} \mathrm{cm}$,
and
\begin{eqnarray}
  \Delta t \sim 10^2 t_p \sim 10^{-25} \,\mathrm{s},
\end{eqnarray}
which is in principle testable at current or future gravity--wave
interferometers, provided there is a controlled way to
distinguish this effect from conventional noise sources~\footnote{In 
new inteferometers the length travelled by the photons 
may actually be higher, of order $10^{5}~{\rm cm}$, and one
should probably 
use the optical length of the photons for such tests, 
which may be higher than the actual length by a 
factor of order $10^2-10^3$.
We thank G. Amelino-Camelia for pointing this out.}.
However, it should be stressed
that this estimate is very preliminary and much more 
detailed work is necessary before even tentative conclusions 
are reached regarding experimental tests 
of the effect in inteferometers.
The above-mentioned sensitivity estimates could become
meaningful only when a detailed description
of the time of observation and the inteferometer spectrum
is given.

Before closing this section, we comment briefly on the
computation of the same phenomenon in extra dimensions
made in the last paper of~\cite{ford}. 
Due to the periodic boundary conditions imposed in that case, there
are Kaluza--Klein modes which are resummed using the image method.
The case of more than one extra dimension complicates the analysis and
was not considered in detail in~\cite{ford}.  For one extra
dimension, the light--cone broadening effect was found to scale
linearly with the distance travelled by the photon, in contrast to our
estimate (\ref{twothirteen}).  We stress therefore that this scaling
depends crucially on the boundary conditions as well as the number of
non--compact dimensions.

\section{Conclusions} 

We have discussed here some possible low-energy probes of
quantum gravity, concentrating on
the possibility that the velocity of light might
depend on its frequency, i.e., the corresponding photon energy.
This idea is very speculative, and the model calculations that we have
reviewed require justification and refinement. However, we
feel that the suggestion is well motivated by the basic fact
that gravity abhors rigid bodies, and the
related intuition that the vacuum should exhibit back-reaction
effects and act as a non-trivial medium. We recall that these
features have appeared in several approaches to quantum gravity,
including the canonical approach and ideas based on extra dimensions.
Therefore, we consider
the motivation from fundamental physics for a frequency-dependent
velocity of light, and the potential significance 
of any possible observation, to be sufficient to examine
this possibility from a phenomenological point of view.

As could be expected, we have found no significant effect in the data
available on GRBs~\cite{efmmn}, 
either in the possible delay times of photons of
higher energies, or in the possible stochastic spreads of
velocities of photons with the same energy. 
However, it has been established that such probes may be
sensitive to scales approaching the Planck mass, if these
effects are linear in the photon energy.
We expect that the redshifts of many more GRBs will become
known in the near future, as alerts and follow-up
observations become more effective, for example
after the launch of the HETE~II satellite~\cite{HETE,Hurley}.
Observations of higher-energy photons from GRBs would be
very valuable, since they would provide a longer
lever arm in the search for energy-dependent effects on photon
propagation. Such higher-energy observations
could be provided by future space experiments such as AMS~\cite{AMS} and
GLAST~\cite{GLAST}.

This is not the only way in which quantum gravity might be probed:
an example that we have advertized previously is provided by
tests of quantum mechanics in the neutral-kaon
system~\cite{ellis84,ELMN,emncplear}. Forthcoming
data from the DA$\Phi$NE accelerator may provide new opportunities for
this quest.
An alternative possibility might be provided by
inteferometric devices intended to detect gravity waves~\cite{amelino99}.
We also regard the emerging astrophysical suggestion of non-vanishing
cosmological vacuum energy as a great opportunity for theoretical
physics. If confirmed, this would provide a number to calculate in a
complete quantum theory of gravity. We have emphasized here
the possibility that this vacuum energy might not be constant, but might
actually be relaxing towards zero, a possibility that may be
tested by forthcoming cosmological observations. We therefore
believe that the phrase `experimental quantum gravity' may not
be an oxymoron.

\section*{Acknowledgements} 

This work is based on invited talks by N.E.M. and D.V.N. 
at the meeting {\it Beyond the Desert 99}, Castle 
Ringberg, Germany, June 6-12 1999. 
We thank Prof. Klapdor-Kleingrothaus and the 
other organizers of this stimulating meeting 
for their interest in our work. 
We also wish to thank G. Amelino-Camelia, I. Antoniadis, 
A. Campbell-Smith, S. Sarkar
and E. Winstanley for discussions. 
The work of N.E.M. is partially supported by P.P.A.R.C. (U.K.).
That of D.V.N. is supported in part by D.O.E. grant DE-F-G03-95-ER-40917.

\end{document}